\documentstyle[11pt,fleqn]{article}
\title{Comments on "Attainable conditions and exact invariant for the
time-dependent harmonic oscillator"}
\author{Kwang-Hua W.  Chu} %\thanks{This paper has not been
\date{P.O. Box 39, Tou-Di-Ban, Road XiHong, Urumqi 830000, PR China}
\begin{document}           % End of preamble and beginning of text.
%%\renewcommand{\baselinestretch}{1}
%\baselinestretch=1        %\intextskip=6mm
\maketitle                 % Produces the title.
\begin{abstract}
We make remarks on Fern\'{a}ndez Guasti's paper [{\it J. Phys. A:
Math. Gen.} 39 (2006) 11825-11832] by pointing out some mistakes
Fern\'{a}ndez Guasti derived therein. \newline

\end{abstract}
%----------------------------------------------------------------------
\doublerulesep=6mm        %\parskip=12 pt
\baselineskip=6mm
\oddsidemargin+1mm         %%\evensidemargin-12mm
\bibliographystyle{plain}
Fern\'{a}ndez Guasti just showed that a time-dependent oscillator equation could be solved numerically for various
trajectories in amplitude and phase variables [1]. His  solutions exhibit a finite
time-dependent parameter whenever the squared amplitude times the derivative
of the phase is invariant. Most of his results were based on the second-order linear
non-autonomous differential equation
\begin{equation}
 \frac{d^2 q(t)}{d\, t^2}+ \Omega^2(t) q(t)=0
\end{equation}
which might be related to a Schr\"{o}dinger equation [2]. \newline
Let the coordinate variable $q$ be written in terms of amplitude $A$ and phase $s$ variables as
\begin{equation}
  q(t) = A(t) cos[s(t)].
\end{equation}
Thus, the equation (1) can also be written as
\begin{equation}
 \Omega^2 (t)=-\frac{\frac{d^2 \{A(t) cos[s(t)]\}}{d\, t^2}}{A(t) cos[s(t)]}.
\end{equation}
Fern\'{a}ndez Guasti then presented some simplified examples together with  the mathematical expression of
$\Omega^2 (t)$ (figures included).  \newline
The present author followed Fern\'{a}ndez Guasti's approaches, but got different results (say,
the mathematical expression of $\Omega^2 (t)$ for Example 1 (cf. Attenuated amplitude in [1]).
To be systematic and specific, some remarks could be made below for the readers' interests
or comparison.  \newline
Firstly, equations (2) and (3) could give us
\begin{equation}
 {\frac{d \{A(t) cos[s(t)]\}}{d\, t}}=\dot{A}\, \cos[s(t)]-A \,\dot{s}\,
 \sin[s(t)],
\end{equation}
\begin{equation}
 {\frac{d^2 \{A(t) cos[s(t)]\}}{d\, t^2}}=\ddot{A}\, \cos[s(t)]-
 2 \dot{A} \,\dot{s}\, \sin[s(t)]-A\, \ddot{s}\,\sin[s(t)] -{A}\,(\dot{s})^2 \,
  \cos[s(t)],
\end{equation}
where $\dot{A}={d A(t) }/{d t}$, $\dot{s}={d s(t) }/{d t}$,
$\ddot{A}={d^2 A(t) }/{d t^2}$, $\ddot{s}={d^2 s(t) }/{d t^2}$.
Based on these expressions, we can easily derive $\Omega^2 (t)$
following the equation (3). For the illustrated Example 1 (by
Fern\'{a}ndez Guasti in [1]), as $A(t) = A_0\sqrt{2}\,[3+
\tanh(\alpha\,t)]^2$, $s(t)=\omega\, t$, ($\omega$ is the
frequency and is a constant), we found out one mistake (different
sign) presented in [1] : the sign of $\alpha \tanh(\alpha\,t)$
inside the second term :[$\omega \tan(\omega\,t)+\alpha
\tanh(\alpha\,t)$]
 of the numerator for $\Omega^2 (t)$
should be negative ($-$) instead of positive ($+$) (i.e.,
it should be read as [$\omega \tan(\omega\,t)-\alpha \tanh(\alpha\,t)$]
inside the second term of the numerator)!
This is understandable considering the first and second terms in the right-hand-side
(RHS) of the equation (5)
above.
\newline
Meanwhile Fern\'{a}ndez Guasti missed the consideration of
$\cos[s(t)]=0$ ($s(t)=(n\pm 1/2)\pi, n=0,\pm1,\pm2, \cdots$) which will cause
singularities or divergences in the (RHS) of the equation (3) or
$\Omega^2 (t)$. That's why there are many kinks for
those curves relevant to $\Omega (t)$ shown in Figs. 1 and 2
(cf. red upper curves in [1] therein).  \newline
Furthermore, Fern\'{a}ndez Guasti demonstrated the last example in the Section ({\it Exact Invariant})
with the prescribed (cf. the equation (7) in [1])
\begin{equation}
  s=\frac{\omega\{ 3\alpha t+\log[\cosh(\alpha t)]\}}{2 \alpha},
\end{equation}
and $A(t) = A_0\sqrt{2}\,[3+ \tanh(\alpha\,t)]^2$.  He obtained
a complicated expression of $\Omega (t)$ (cf. the long mathematical expression below the equation (7) in [1]
which is at page 11829 of [1]) which after the present author checked is also false! In fact,
we derived
\begin{equation}
 \Omega (t) = \frac{\omega^2}{4}[3+\tanh(\alpha t)]^2+\frac{\alpha^2\, \tanh(\alpha t)}{\cosh^2 (\alpha t)
   [3+\tanh(\alpha t)]}-\frac{3}{4}\frac{\alpha^2}{\cosh^4 (\alpha t)[3+\tanh(\alpha t)]^2},
\end{equation}
by noting that, from our equation (5), for the prescribed $A(t)$
and $s(t)$, $A\, \ddot{s}=-2 \dot{A} \,\dot{s}$ which can simplify
many mathematical manipulations (the present author thought that
Fern\'{a}ndez Guasti possibly mishandled the signs of both $A\,
\ddot{s}$ and $2 \dot{A} \,\dot{s}$ and then got the false
expression in [1]!). Here, we remind the readers that
\begin{displaymath}
  \dot{s}=\frac{\omega[3+\tanh(\alpha t)]}{2}, \hspace*{12mm}
  \ddot{s}=\frac{\omega \alpha}{2 \cosh^2 (\alpha t)}.
\end{displaymath}
 {\it Acknowledgements.} The author is
partially supported by the Starting Funds of 2005-XJU-Scholars.
\newline


\vspace*{10mm}

\begin{thebibliography}{99}
\doublerulesep=2mm        %\parskip=12 pt
\baselineskip=3mm
\bibitem{S:Plate}  Fern\'{a}ndez Guasti M 2006 {\it J. Phys. A: Math. Gen.} {\bf 39}  11825.  %¨C11832
\bibitem{ED:Continuum} Bender CM  and Orszag SA 1978 {\it Advanced Mathematical Methods for
Scientists and Engineers} (McGraw-Hill, Inc., New York) pp. 486.
%Landau LD, Lifshitz EM and Pitaevskii LP 1984 {\it Electrodynamics of Continuous Media} %(2nd edition, Pergamon, Oxford).
%Israelachvili JN 2000 {\it Intermolecular and Surface Forces} (2nd edition, Academic Press, San Diego).
\end{thebibliography}
\end{document}